\documentclass[12pt,a4paper]{article}

\usepackage{latexsym}
\usepackage{epsf}
\usepackage{graphicx}
\def\be{\begin{equation}}
\def\ee{\end{equation}}
\def\bea{\begin{eqnarray}}
\def\eea{\end{eqnarray}}
\def\by{\left(\begin{array}}
\def\ey{\end{array}\right)}

\def\slash#1{\setbox0=\hbox{$#1$}#1\hskip-\wd0\dimen0=5pt\advance
       \dimen0 by-\ht0\advance\dimen0 by\dp0\lower0.5\dimen0\hbox
         to\wd0{\hss\sl/\/\hss}}

\newcommand{\bl}{{\bf LEFT}}
\newcommand{\br}{{\bf RIGHT}}

\begin{document}

\begin{center}
\Large{Freeze Out Process with In-Medium Nucleon Mass} 
\end{center}

\begin{center}

\small {Sven Zschocke$^{1,2}$, Laszlo P. Csernai$^{1,3}$, 
Etele Moln\'ar$^{1}$, Jaakko Manninen$^{1,4}$, Agnes Nyiri$^{1}$} 
\end{center}

\footnotesize{
\begin{center}
$^{1}$ Section for Theoretical and Computational Physics, and
Bergen Computational Physics Laboratory, \\
University of Bergen, 5007 Bergen, Norway \\
\vspace{0.2cm}
$^{2}$ Forschungszentrum Rossendorf, 01314 Dresden, Germany\\
\vspace{0.2cm}
$^{3}$ MTA-KFKI, Research Institute of Particle and Nuclear Physics,\\ 
1525 Budapest 114, Hungary\\
\vspace{0.2cm}
$^{4}$ University of Oulu, Department of Physical Sciences, 90571 Oulu, 
Finland
\vspace{0.2cm}
\end{center}
}
\normalsize
\date{\today}

\begin{abstract}
We investigate the kinetic freeze out scenario of a 
nucleon gas through a finite layer. The in-medium mass 
modification of nucleons and it's impact on the freeze 
out process is studied. A considerable modification of 
the thermodynamical parameters temperature, flow-velocity, 
energy density and particle density has been found 
in comparison with evaluations which use a constant 
vacuum nucleon mass.  
\end{abstract}

\section{Introduction}\label{intro}

As the fireball of a nucleus-nucleus collision cools down  
below a certain freeze out temperature $T_{FO} \le T_c \simeq 175$ MeV 
the inelastic (chemical freeze out) and elastic (kinetic freeze out) 
collisions among the hadrons cease.  
This process is usually called the freeze out scenario (FO).
Several approaches have been applied for the description of the 
freeze out of strongly interacting matter. 
Especially, there are kinetic 
models~\cite{kineticmodels_5, kineticmodels_10} 
as well as hydrodynamical approaches~\cite{hydromodels_5} which have been 
proven to be able to describe most of the collective phenomena like the 
different flow components in heavy-ion reactions. 
Despite the success in comparison with experiments, the 
in-medium modifications of the hadrons during the freeze out process 
have not been taken into account yet. In all of the former evaluations 
the vacuum parameters of the particles have been implemented. 
However, during the freeze out process, the temperature and particle 
densities are
presumably close to the deconfinement phase transition critical values
\cite{chemical_10}. Accordingly, strong in-medium modifications of hadronic 
properties like mass, width, coupling constants etc., are expected. 
The question arises how strong their impact on the freeze out process is. 
For answering this issue we study a nucleon gas 
and investigate how strong the impact of an in-medium nucleon mass shift 
on the thermodynamical parameters, temperature, flow-velocity,
energy density and particle density, of the freeze out scenario is.   

\section{Freeze Out Process within a finite layer}\label{FO}

The theoretical description of the kinetic freeze out
within a hydrodynamical approach has been worked out
some years ago; see \cite{finitelayer_10} and references therein. 
Here, the evaluations presented are based on the approach of 
Ref. \cite{timelike_5}, where the freeze out description has been 
generalized to the case of a finite time-like layer 
(a finite space-like layer has also been investigated in  
\cite{spacelike_5}; more details about the approach presented 
will be published in \cite{Bergen}). 
Accordingly, 
local equilibrium implies that the thermodynamical parameters inside the layer
become space-time dependent, i.e. we have a space-time dependent temperature
$T(x)$, flow velocity $v(x)$, energy density $e(x)$ and
nucleon density $n(x)$.
From the equations of hydrodynamics we derive the following 
set of coupled differential equations for three unknowns  
\bea
d e (x) &=& u_{\mu} (x) \, d T^{\mu \nu} (x) \, u_{\nu} (x) \,
+ \, 2 \, d u_{\mu} (x) \, T^{\mu \nu} (x) \, u_{\nu}\;,
\label{hydro_35}
\\
d n (x) &=& u_{\mu} (x) \, d N^{\mu} (x) \;,
\label{hydro_40}
\\
d u^{\mu} (x) &=& \frac{1}{n (x)} \;
\bigg( g^{\mu \nu} - u^{\mu} (x) u^{\nu} (x) \bigg) \; d N_{\nu}\;.
\label{hydro_45}
\eea
The differentials in (\ref{hydro_35}) - (\ref{hydro_45})
are deduced from the microscopic definition,  
$d\, N^{\mu} (x) = \int \frac{d^3 {\bf k}}{k^0} k^{\mu} \;d f(x, k)$
for the particle current, and
$d\,T^{\mu \nu} (x) = \int \frac{d^3 {\bf k}}{k^0} k^{\mu} k^{\nu}\; d f(x, k)$
for the energy momentum tensor. Here,
$x^{\mu} = (t, {\bf r})$ is the four-coordinate and
$k^{\mu} = (E_k, {\bf k})$ is the four-momentum of the nucleon;
the scalar one-particle distribution function is 
normalized to the invariant number of nucleons $N$,
i.e. $N = \int d^3 {\bf r} \; d^3 {\bf k} \; f (x, k)$.
We mention that the second term in (\ref{hydro_35}) vanishes
within the approach presented, but not in general.
The Rest Frame of the Gas (RFG) is defined by
$u^{\mu}_{\rm RFG} (x) = (1, 0, 0, 0)$. For baryon-dominated matter, 
like in our case, the Eckart's definition 
\cite{book_1} of four-flow is commonly used, defined by
$u^{\mu} (x) = N^{\mu} (x)\,/\,\sqrt{N^{\nu} (x) N_{\nu} (x)}$.
Another Lorentz frame, e.g. Rest Frame of the Front (RFF), 
can be defined by a Lorentz boost in respect to RFG.

Since there are four unknowns in the problem under consideration 
an additional constraint is necessary, which is provided by the 
Equation of State (EOS)
for the nucleon gas \cite{EOS_5}
\bea
e (x) = n (x) \bigg[M_N (n(x) ,T(x)) \, - \,  E_0 \, + \, \frac{K}{18} \,
\left(\frac{n (x)}{n_0} - 1 \right)^2 \, + \, \frac{3}{2} \; T (x)  \bigg] \;.
\label{eos_5}
\eea
The nuclear binding energy is $E_0 = 16$ MeV, and  
$K =\simeq 235$ MeV is the compressibility; $M_N (n,T)$ is the 
in-medium nucleon pole mass. 
The EOS (\ref{eos_5}) is used to determine the temperature $T (x)$ of
the interacting component of 
the nucleon gas during the freeze out process. Accordingly, the
four equations (\ref{hydro_35}), (\ref{hydro_40}), (\ref{hydro_45}) and
(\ref{eos_5}) represent a closed set for evaluating the four unknowns
$T, v, e, n$ of the one-particle system.

Furthermore, as the system expands and cools down the number of interacting
particles decreases up to the post freeze out surface of the finite layer, 
where by definition the density of interacting particles vanishes.
Correspondingly,   
the particle distribution function is decomposed into two components, an  
interacting part $f_i$ and a non-interacting part $f_f$, thus
$f (x, k) = f_i (x, k) \; + \; f_f (x, k)$.
Accordingly, there is an interacting particle density 
$n_i$ and a non-interacting particle density $n_f$
with $n = n_f + n_i$. 
On the pre-freeze out hypersurface we assume to have thermal equilibrium, i.e. 
we have a J\"uttner distribution, cf. \cite{book_1}, 
for $f_i$ as starting one-particle 
distribution function, while by definition $f_f$ 
is zero on the pre-freeze out 
hyper-surface.  
The space-time evolution of the interacting and non-interacting components 
inside the layer is governed by the following differential equations 
\cite{timelike_5}:
\bea
\partial_t \, f_i &=& - \frac{1}{\tau} \left( \frac{L}{L - t} \right) 
\left(\frac{k^{\mu} \, d \sigma_{\mu}}{k_{\mu} \, u^{\mu}} \right) f_i 
\; + \; \frac{1}{\tau_0} [f_{eq} (t) - f_i] \;,
\label{kinetic_10}
\\
\nonumber\\
\partial_t \, f_f &=& + \frac{1}{\tau} \left( \frac{L}{L - t} \right) 
\left(\frac{k^{\mu} \, d \sigma_{\mu}}{k_{\mu} \, u^{\mu}} \right) f_i \;, 
\label{kinetic_15}
\eea
with the time $\tau$ between collisions, and $f_{eq}$ is  
the J\"uttner distribution, cf. \cite{book_1}. 
The second term in (\ref{kinetic_10}) is the re-thermalization term  
\cite{finitelayer_10}, which describes how fast the interacting 
component approaches the J\"uttner distribution within a relaxation time 
$\tau_0$. Here, we will use the immediate re-thermalization limit 
$\tau_0 \rightarrow 0$, which implies $f_i \rightarrow f_{eq}$ 
faster than $\tau_0 \rightarrow 0$.
The explicit expressions for the differentials 
$d\,N^{\mu}$ and $d\, T^{\mu \nu}$ within the approach presented can be found
in \cite{timelike_5}. 

The set of equations (\ref{hydro_35}) - (\ref{kinetic_15})  
allow us to  
evaluate the basic thermodynamical function $T(x), v(x), e(x)$ 
and $n(x)$ during the freeze out process for a particle with a 
mass $M_N (n,T)$. 

\section{Nucleon Mass Shift}\label{mass_shift}

A generalization of the QCD sum rule approach for nucleons at 
finite densities and 
temperatures \cite{nucleonsumrule_10} leads to  
the following expression for the pole mass of a nucleon  
embedded in a hot and dense hadronic medium,
\bea
M_N (n, T) &=& M_N (0) + {\rm Re}\, \Sigma_S (n, T) +
{\rm Re}\, \Sigma_V (n, T)\;,
\label{massshift_3}
\eea
where $M_N(0) = 939$ MeV is the vacuum nucleon pole mass, and
with the attractive scalar part (${\rm Re}\; \Sigma_S < 0$) and  
the repulsive vector part (${\rm Re}\; \Sigma_V > 0$) of nucleon self energy 
in medium.  
Typical values at nuclear saturation density 
$n_0 = 0.17\, {\rm fm}^{-3}$ 
and at vanishing temperature 
are ${\rm Re}\; \Sigma_S = - 400$ MeV,
${\rm Re}\;\Sigma_V = + 300$ MeV 
\cite{nucleonsumrule_10}.  
Here, we will take the QCD sum rule results for a nucleon 
in matter, given by \cite{nucleonsumrule_10} 
\bea
{\rm Re}\;\Sigma_S (n, T) &=& + M_N (0) 
\left( \frac{\langle \Omega | {\overline q} q | \Omega \rangle} 
{\langle 0 | {\overline q} q | 0 \rangle} - 1 \right) \;, 
\label{massshift_4}
\\
\nonumber\\
{\rm Re}\;\Sigma_V (n, T) &=& - \frac{8}{3} M_N (0) 
\frac{\langle \Omega | q^{\dagger} q | \Omega \rangle}
{\langle 0 | {\overline q} q | 0 \rangle} \;.
\label{massshift_5}
\eea
Here, $\langle \Omega | {\overline q} q | \Omega \rangle$ and 
$\langle \Omega | q^{\dagger} q | \Omega \rangle$ are in-medium condensates 
and $|\Omega\rangle$ is a state which describes 
the hot and dense hadronic matter 
inside the layer, while $\langle 0| {\overline q} q |0\rangle
= (-0.250\;{\rm GeV})^3$ is the chiral condensate. 
In (\ref{massshift_4}) and (\ref{massshift_5}) we have neglected 
the gluon condensate and higher mass 
dimension condensates which give rise to small corrections only. 

There are two nucleonic components inside the finite layer: 
an interacting component with density $n_i$ and a non-interacting  
component with density $n_f$. For evaluating the condensates 
(\ref{massshift_4}) and (\ref{massshift_5}) 
we approximate the interacting component 
by a Fermi gas with chemical potential $\mu_i$ and temperature $T$. 
On the other side, the temperature for the non-interacting component 
becomes ill-defined. Nonetheless, a relevant physical parameter for describing 
the non-interacting component remains the density $n_f$. 
Accordingly, the condensates in one-particle approximation 
are given as follows \cite{change_condensate2}:  
\bea
\langle \Omega | {\overline q} q | \Omega \rangle &=& 
\langle 0 | {\overline q} q | 0 \rangle \, + \, \hat{I} (\mu_i,T) \;
\langle N ({\bf k}) | {\overline q} q | N ({\bf k}) \rangle\; 
+ \frac{n_f}{2 M_N(0)} \langle N ({\bf k}) | {\overline q} q | N ({\bf k}) 
\rangle \;,  
\label{condensates_5}
\\
\nonumber\\
\langle \Omega | q^{\dagger} q | \Omega \rangle &=& 
\hat{I} (\mu_i,T)\; \langle N ({\bf k}) | q^{\dagger} q | N ({\bf k}) \rangle\;
+ \frac{n_f}{2 M_N(0)} \langle N ({\bf k}) | 
q^{\dagger} q | N ({\bf k}) \rangle \;,
\label{condensates_6}
\eea
with $\hat{I} \equiv 4\int\,d^3{\bf k}/[(2 E_k)\, (2 \pi)^3)\,
({\rm exp}((E_k-\mu_i)/T)+1)]$; for vanishing temperature we have  
$\hat{I} \rightarrow n_i/(2 M_N)$. 
Note that $\langle 0 |q^{\dagger} q | 0 \rangle = 0$;  
the nucleon energy is $E_k = \sqrt{M_N(0)^2 + {\bf k}^2}$.
For nucleons we take the relativistic normalization 
$\langle N ({\bf k}_1) |
N ({\bf k}_2) \rangle = 2 E_{k_1} (2 \pi)^3 \delta^{(3)}
({\bf k}_1 - {\bf k}_2)$ is used.
The chemical potential for the interacting component can be
evaluated via
$n_i = 4 \int d^3 {\bf k}/[(2 \pi)^3\,({\rm exp}((E_k-\mu_i)/T)+1)]$. 
The condensates in Fermi gas approximation are given by \cite{sigmaterm_5}
$\langle N ({\bf k}) | {\overline q} q | N ({\bf k}) \rangle
= M_N (0) \sigma_N/m_q$ and  
$\langle N ({\bf k}) | q^{\dagger} q | N ({\bf k}) \rangle
= 3 M_N (0)$. 
The nucleon sigma term is $\sigma_N \simeq 50$ MeV,   
and $m_q \simeq 5$ MeV is the averaged current quark mass 
of the light quarks.  
Inserting these parameters into (\ref{massshift_4}) and 
(\ref{massshift_5}) we obtain 
${\rm Re} \; \Sigma_S = - 390$ MeV and ${\rm Re}\;\Sigma_V = 315$ MeV at 
ground state saturation density $n_0$. 
The Eqn.~(\ref{massshift_3}) - (\ref{condensates_6})  
summarize our propositions made for obtaining the  
in-medium nucleon pole mass $M_N (n, T)$, which enters the EOS (\ref{eos_5}) 
and the differentials $d N^{\mu}$ and $d T^{\mu \nu}$.

\section{Results and Discussion}\label{results_discussions}

For all of the calculations, we have taken 
$T_{{\rm pre}\;{\rm FO}} = 150$ MeV, 
$n_{{\rm pre}\;{\rm FO}} = 1.5 \, n_0$
and $v_{{\rm pre}\;{\rm FO}} = 0.5 \,c$
as starting values on the pre freeze out hypersurface.
These values are, for instance, in line with typical parameters 
which have been reached within the Alternating-Gradient Synchrotron 
(AGS) at {\it Brookhaven National Laboratory} (BNL) in Brookhaven/USA. 
Higher baryonic densities can be reached within the  
Schwer-Ionen-Synchrotron (SIS) at 
{\it Gesellschaft f\"ur Schwerionenforschung} (GSI) 
in Darmstadt/Germany.  
Note that $T_{{\rm pre}\;{\rm FO}}$ and $n_{{\rm pre}\;{\rm FO}}$ 
are pre freeze out values and, 
therefore, they are larger than typical post freeze out values 
given, for instance, in Ref. \cite{chemical_10}. 

\begin{center}
\begin{figure}[!ht]
\hspace{-0.5cm}\includegraphics[angle=0,scale=0.8]{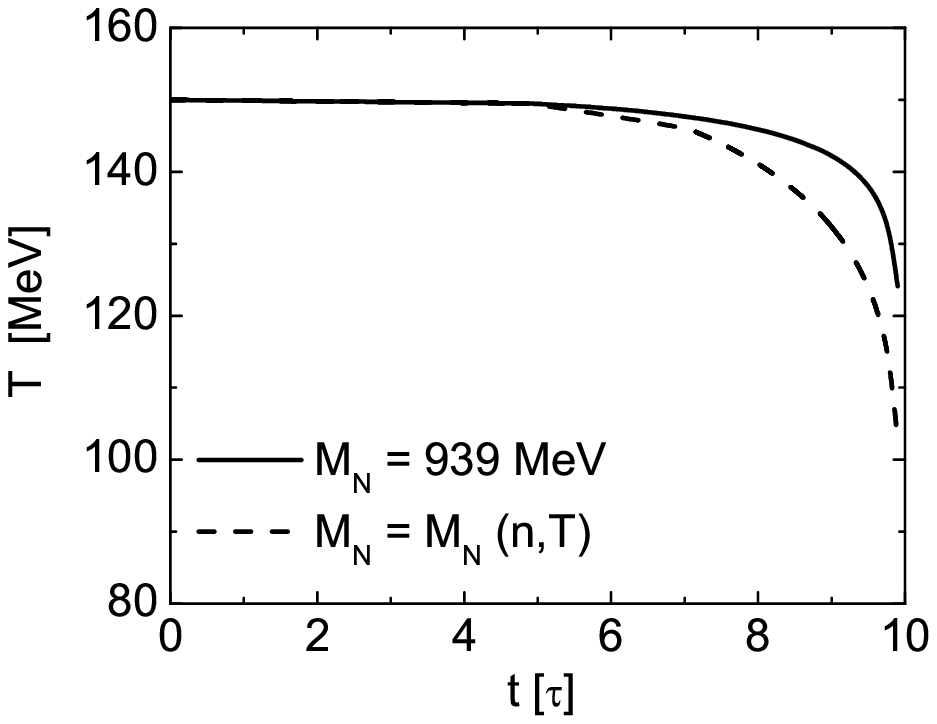}
\hspace{-0.5cm}\includegraphics[angle=0,scale=0.8]{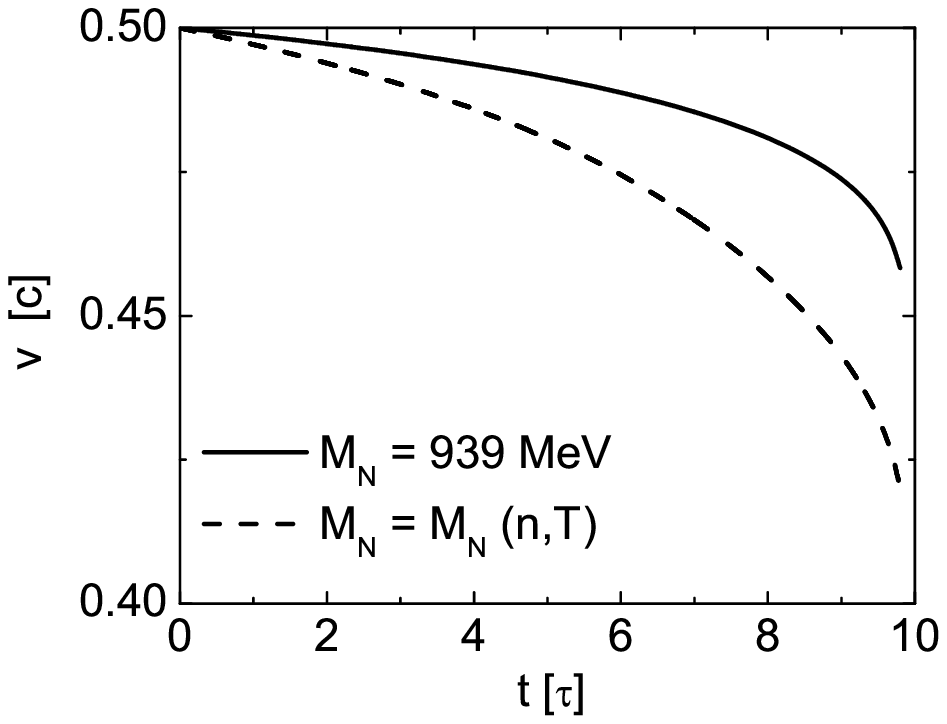}
\caption{
\bl: The temperature of the interacting component. 
\br: The flow velocity parameter $v$ of the interacting component.
The solid lines are with a constant nucleon mass $M_N (0) = 939$ MeV, while 
the dashed curves are evaluated with a density and temperature dependent  
nucleon mass $M_N (n,T)$.\label{fig:tv}} 
\end{figure}
\end{center}

\begin{center}
\begin{figure}[!h]
\hspace{-0.7cm}\includegraphics[angle=0,scale=0.8]{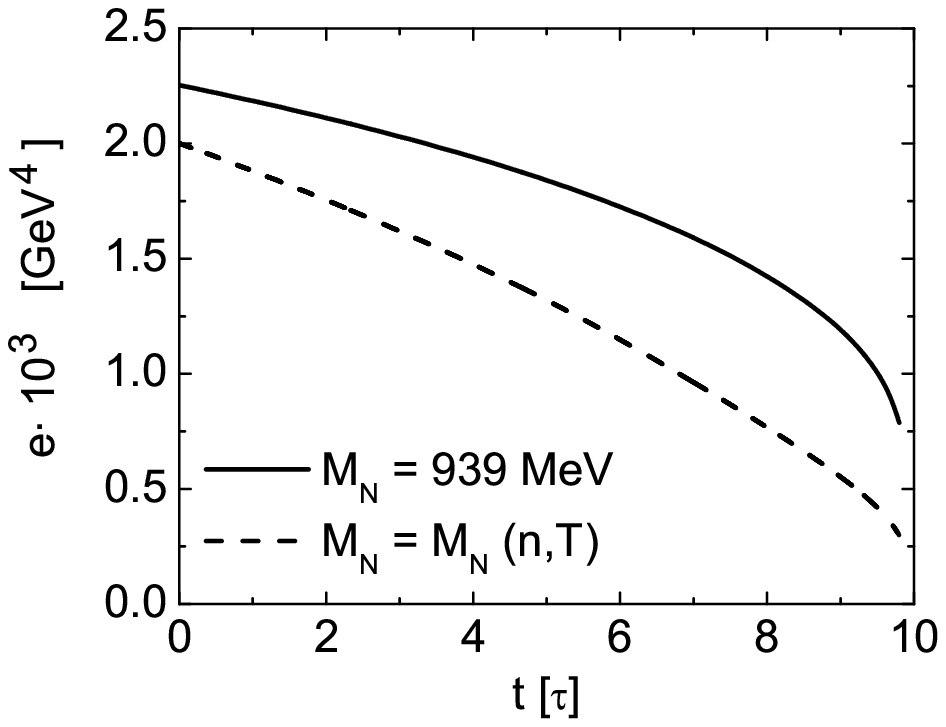}
\hspace{-0.7cm}\includegraphics[angle=0,scale=0.8]{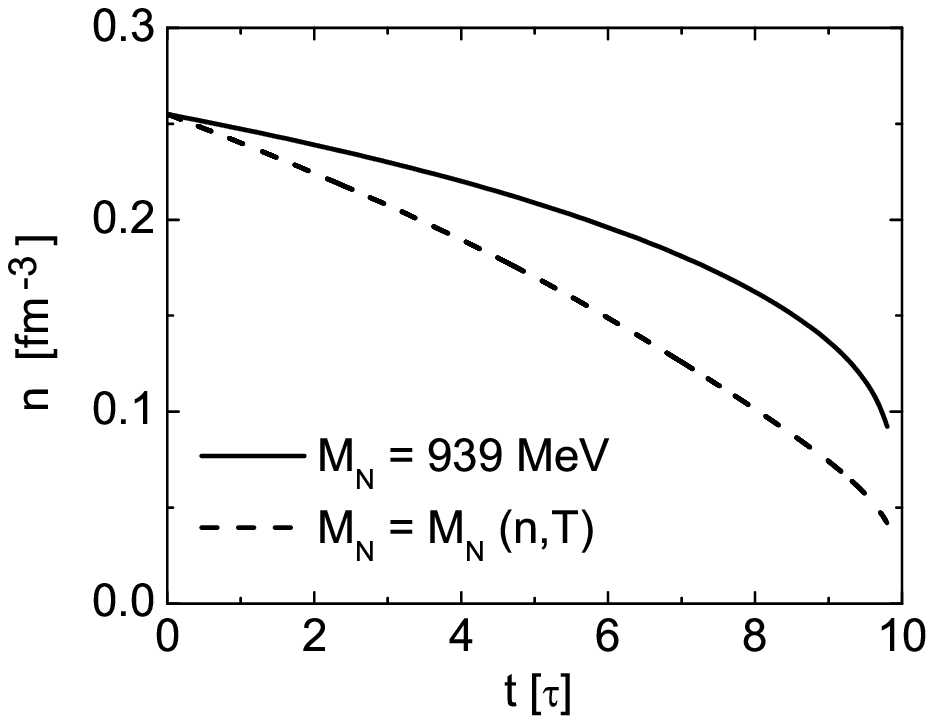}
\caption{
\bl: Nucleon energy density of the interacting component.
\br: Nucleon particle density of the interacting component. 
The solid lines are with a constant nucleon mass $M_N(0) = 939$ MeV, while 
the dashed curves are evaluated with a density and temperature  
dependent nucleon mass $M_N (n,T)$. \label{fig:en} } 
\end{figure}
\end{center}

In Figs.~\ref{fig:tv} and \ref{fig:en}, the time evolution of the primary 
thermodynamical functions through the finite freeze out layer are shown,  
in terms of the proper time $\tau$. Note that the densities    
$n=n_i+n_f$ and $e=e_i+e_f$ are kept constant inside the layer. 

We find a substantial impact of in-medium mass modification
on the freeze out process within the purely nucleon gas model.
Furthermore, the Figs.~\ref{fig:tv} and \ref{fig:en}
also elucidate, that the freeze out process
proceeeds faster for all thermodynamical quantities $T, v, e, n$
when taking into account the mass dropping of nucleons.
The physical reason for a faster freeze out originates from a smaller
energy density of the nucleon system due to a smaller nucleon mass
$M_N(n,T)$ compared to the vacuum nucleon mass $M_N(0)$.

Finally, we remark that in-medium modifications have actually    
to be taken into account not only during the freeze out process, but 
also before, i.e. during the hadronization. 
This points then to an even 
stronger impact of in-medium modifications on the final particle
spectrum than presented. 

\section{Summary}\label{sum}

We have investigated a freeze out scenario within a finite layer 
for a massive nucleon gas. Special attention has been drawn about the issue   
how strong the impact of the in-medium nucleon mass modification on the 
thermal freeze out process is. 
By focussing on a purely nucleon gas we have found a  
substantial effect 
on the thermodynamical quantities like 
temperature $T$, flow velocity $v$, 
particle density $n$ and 
energy density $e$  of the interacting component.  
All of these thermodynamical functions have revealed a faster freeze out 
compared to a scenario without an n-medium nucleon mass shift.  

In summary, our findings for a nucleon gas suggest 
that taking into account in-medium modifications of nucleons 
seems to be necessary and an interesting phenomenon,  
in particular for collision scenarios with high baryonic densities. 

\section*{Acknowledgements}  
The authors would like to express their gratitude to 
Prof. J. Cleymans for enlighting discussions. 
S.Z. and J.M. wishes to acknowledge the NordForsk for the financial 
support of this work. 
S.Z. thanks for the warm hospitality at the Bergen Center for Computational 
Science (BCCS) and Bergen Physics Laboratory (BCPL) at the University of 
Bergen/Norway. 

\end{document}